\documentclass[epj,final]{svjour} 
\usepackage{color}
\usepackage{amsfonts}
\usepackage{amsbsy}
\usepackage{mathrsfs}
\usepackage{graphicx}
\def\lsim{\mathrel{\rlap{
\lower4pt\hbox{\hskip-3pt$\sim$}}
    \raise1pt\hbox{$<$}}}     
\def\gsim{\mathrel{\rlap{
\lower4pt\hbox{\hskip-3pt$\sim$}}
    \raise1pt\hbox{$>$}}}     
\def\scr#1{\mbox{\scriptsize #1}}
\begin{document}
\title{
Entropy Production and Effective Viscosity in Heavy-Ion Collisions} 
\titlerunning{Entropy Production and Effective Viscosity}
\author{Yu. B. Ivanov\inst{1,2}  \thanks{e-mail: Y.Ivanov@gsi.de} \and
A. A. Soldatov\inst{2} \thanks{e-mail: saa@ru.net}}
\institute{%
National Research Centre "Kurchatov Institute", 
123182 Moscow, Russia
\and
National Research Nuclear University "MEPhI" (Moscow Engineering
Physics Institute), 115409 Moscow, Russia}
\date{Received: date / Revised version: date} 

\abstract{
Entropy production and an effective viscosity in central Au+Au collisions  
are estimated  
in a wide range of incident energies
 3.3 GeV  $\le \sqrt{s_{NN}}\le$ 39 GeV. The simulations are performed  
within a three-fluid model 
employing three different equations of state with and without 
deconfinement  transition, which are equally good 
in reproduction of the momentum-integrated elliptic flow of charged particles 
in the considered energy range.  
It is found that 
more than 80\% entropy 
is produced during a short early collision
stage which lasts $\sim$1 fm/c at highest considered energies  
 $\sqrt{s_{NN}}\gsim$ 20 GeV. 
The estimated values of the viscosity-to-entropy ratio ($\eta/s$) are approximately 
the same in all considered scenarios. 
At final stages of the system expansion they range 
from $\sim$0.05 at highest considered energies to $\sim$0.5 lowest ones. 
It is found that the $\eta/s$ ratio decreases with the temperature ($T$) rise 
approximately as $\sim 1/T^4$ and exhibits a rather weak dependence on the 
net-baryon density. 
}
\maketitle

\section{Introduction}

Dissipation in strongly interacting matter is an important property of 
the dynamics of heavy-ion collisions. The entropy is a key quantity that characterizes 
the dissipation. Fast entropy generation at the initial stage of the collision is still one 
of the main challenges for the theory \cite{Berges:2012ks,Fukushima:2016xgg}. The  
low-dissipative collective evolution following this short initial stage is better 
theoretically understood. It is described by hydrodynamics with relatively low 
viscosity. Observables that are the most sensitive to the dissipation at the expansion  
stage of the reaction are elliptic flow and other anisotropic flow coefficients. 
This dissipation deduced from analysis of experimental data at the Large Hadron Collider (LHC) at 
CERN and at top energies of the Relativistic Heavy Ion Collider
(RHIC) at Brookhaven National Laboratory (BNL) amounts to $\eta/s\approx$ 0.1 -- 0.2 
in terms of the viscosity-to-entropy ratio \cite{Heinz:2013th}. Theoretical estimates of this 
ratio in the quark-gluon phase (QGP) are in agreement with the values deduces from the data.

It was expected that at lower collision energies, i.e. at energies of the 
Beam Energy Scan (BES) program at RHIC and below, the viscosity of the matter
should rapidly rise because the system spends most of its time in the hadronic phase
\cite{Kestin:2008bh}. This would result in a reduction of the elliptic flow as 
compared with that at the top RHIC energies. One of the surprises of 
the BES RHIC program  has been that the 
elliptic flow of charged hadrons does
not change significantly when the collision energy is reduced 
from $\sqrt{s_{NN}} =$ 200 to 10 GeV \cite{Adamczyk:2012ku}.
This implied that the $\eta/s$ ratio is not considerably larger at 
low energies than that at high energies. The analysis of the STAR data 
\cite{Adamczyk:2012ku}, performed within a hybrid model \cite{Petersen:2008dd}, 
indeed indicated that the $\eta/s$ ratio increases only up to value of 0.2 at 
the lowest energy of the BES RHIC program, i.e. $\sqrt{s_{NN}} =$ 7.7 GeV \cite{Karpenko:2015xea}. 
In Refs. \cite{Itakura:2007mx,Khvorostukhin:2010aj,Denicol:2013nua,Kadam:2015xsa} it was 
found that a hadronic system with a large net-baryon density is closer to the ideal
fluid limit than the corresponding gas with zero net-baryon density. 
The latter suggests that the system created at lower collision energies may display a fluid-like behavior with an
effective fluidity close to that found at RHIC top-energy collisions, thus explaining why the 
elliptic flow measured at lower RHIC energies  is close to that observed  
at the top RHIC energies.

\begin{figure}[pbt]
\includegraphics[width=6.50cm]{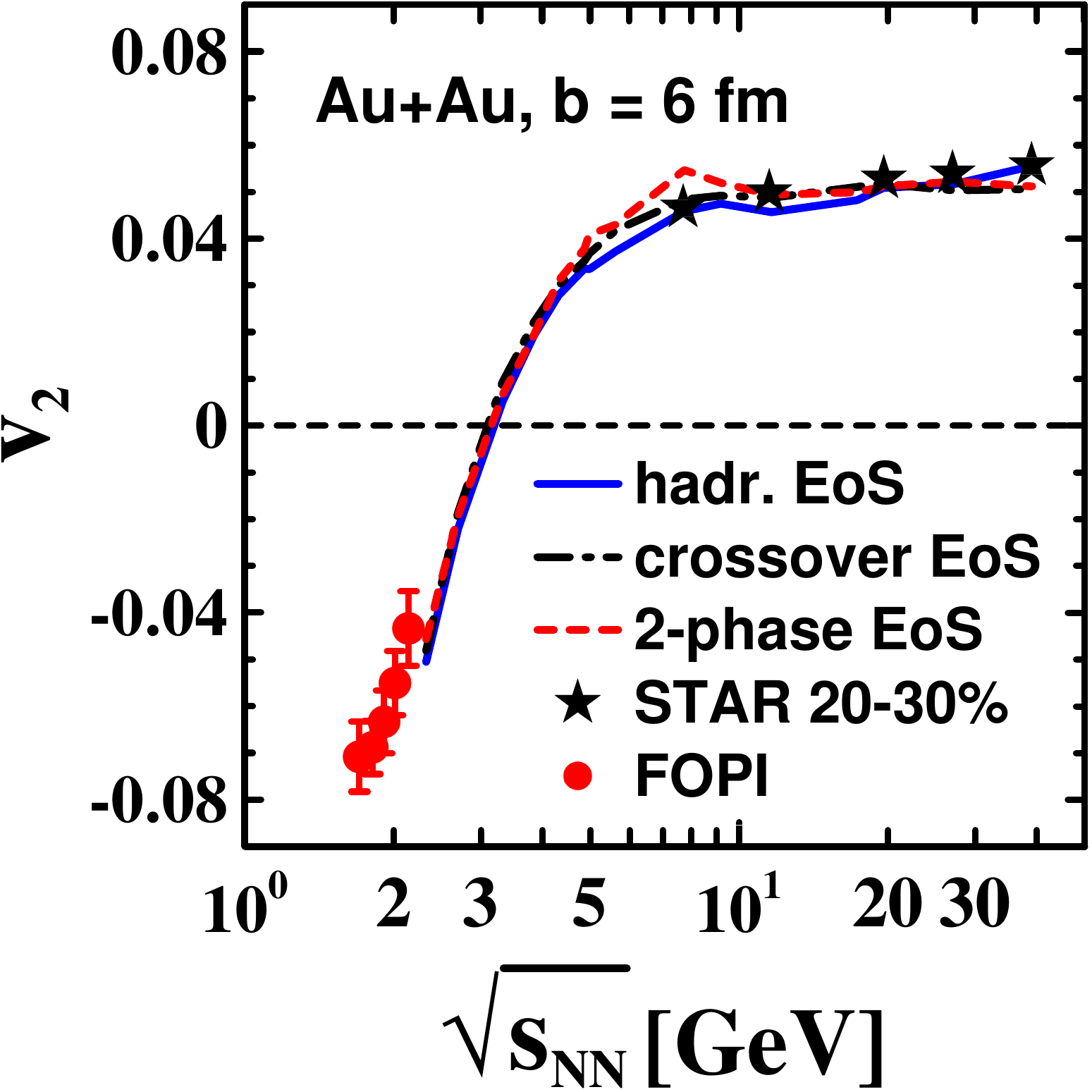}\\
 \caption{
Elliptic flow of charged particles
at midrapidity as a function of incident energy
in mid-central collisions 
Au+Au at impact parameter $b=$ 6 fm. 
Experimental data on the integral elliptic flow of charged particles are from  
STAR  Collaboration  \cite{Adamczyk:2012ku} (subset v2(EP)). 
FOPI data for Z=1 particles \cite{FOPI05}  
are also displayed. 
} 
\label{fig0}
\end{figure}

In our recent paper \cite{Ivanov:2014zqa} we found that the model of the 
three-fluid dynamics (3FD) \cite{3FD} equally well describes the STAR data \cite{Adamczyk:2012ku}
on the transverse-momentum-integrated elliptic flow of charged particles 
at energies from $\sqrt{s_{NN}} =$ 7.7 to 39 GeV within 
very different scenarios characterized by very different 
equations of state (EoS's)---a purely hadronic EoS \cite{gasEOS}  
and two versions of the EoS involving the   deconfinement
 transition \cite{Toneev06}, i.e. a first-order phase transition  
and a smooth crossover one. 
An example of achieved reproduction of the experimental data is presented 
in Fig. \ref{fig0}. 
In Fig. \ref{fig0},  
FOPI data for Z=1 particles \cite{FOPI05}  
are also displayed because Z=1 particles dominate among charged particles
in the respective energy range. 
In Ref. \cite{Ivanov:2014zqa} we failed to answer the question 
why it happens, though we suspected that the reason is that the 
dissipation in the 3FD dynamics with different EoS's is very similar. 
The 3FD model does not include viscosity in its formulation.
However, dissipation is present in the 3FD trough
friction interaction between participated fluids.
Though the 3FD dissipation is not directly associated with the viscosity,
it is desirable to express it in terms of the shear viscosity in order 
to compare it with that in other approaches \cite{Karpenko:2015xea}.

In the present paper we calculate the entropy production in the 3FD simulations 
at various energies and within different scenarios in order to quantify 
the dissipation in the 3FD model. 
To estimate this dissipation in therms of an effective shear viscosity,  
we consider this entropy as if it is generated within the 
conventional one-fluid viscous hydrodynamics. 
This effective  shear viscosity is not a true one characterizing 
the matter near equilibrium. It is just a  shear viscosity that would 
produce the same entropy as that resulting from the nonequilibrium self-diffusion in the 
3FD model. 
The results of this study have been briefly reported in letter \cite{Ivanov:2016vkw}.

\section{The 3FD Model}
\label{Model}

\begin{figure*}[pbt]
\includegraphics[width=16.50cm]{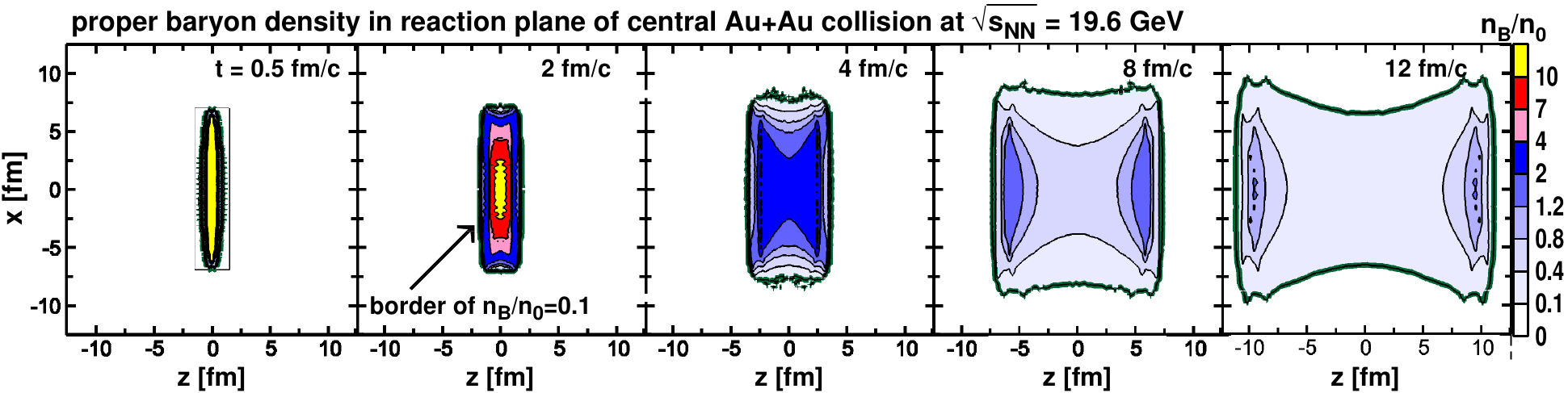}\\
\includegraphics[width=16.50cm]{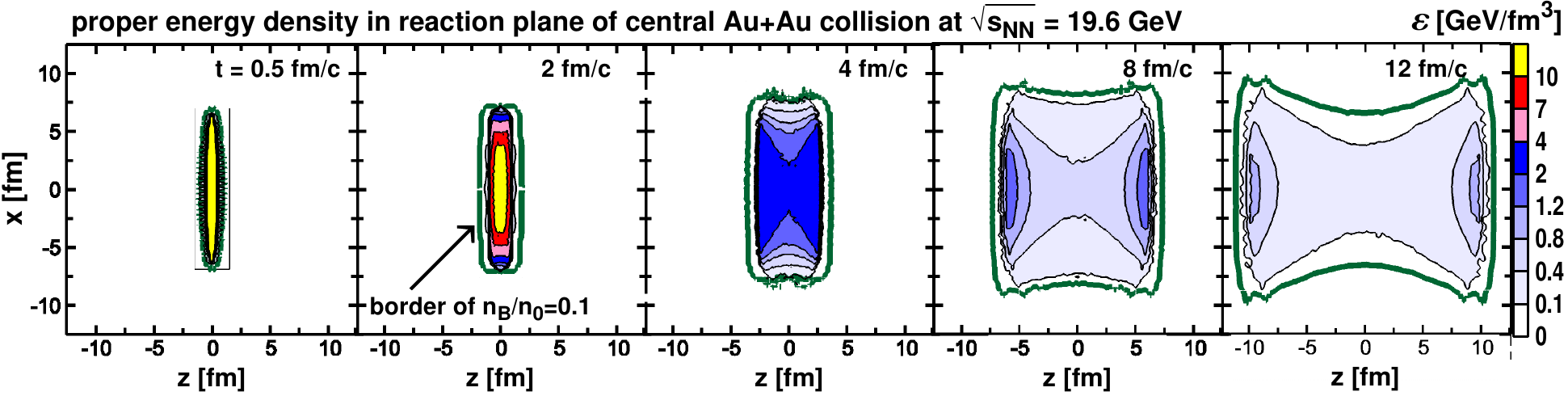}\\
\includegraphics[width=16.50cm]{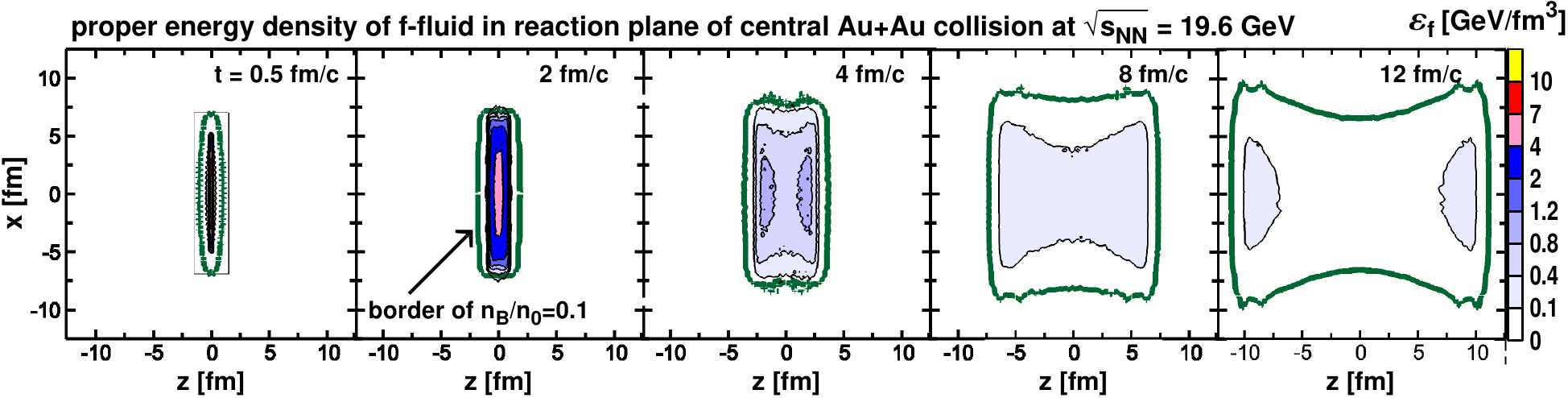}\\
 \caption{
Time evolution of proper baryon density   
$n_B$ in units of normal nuclear density $n_0=$ 0.15 fm$^{-3}$
(upper row of panels), proper energy density 
(middle row of panels) and proper energy density of the f-fluid  
(lower row of panels) in the $xz$ plane (i.e. the reaction plane) 
of central Au+Au collision within the crossover
scenario at $\sqrt{s_{NN}}=$ 19.6 GeV. The bold green contour in 
all the panels confines the region where $n_B/n_0>$ 0.1. 
} 
\label{fig1a}
\end{figure*}

The 3FD model treats \cite{3FD} the collision process
within the fluid dynamics from the very beginning, 
i.e. the stage of cold nuclei, up to freeze-out. The 3-fluid approximation is a minimal way to 
simulate the finite stopping power at the initial stage of the collision.
Within this approximation 
a generally nonequilibrium distribution of baryon-rich
matter is modeled by counter-streaming baryon-rich fluids 
initially associated with constituent nucleons of the projectile
(p) and target (t) nuclei. In addition, newly produced particles,
populating the midrapidity region, are associated with 
a separate net-baryon-free fluid---so called ``fireball'' fluid (f-fluid). 
A certain formation time $\tau_f$ is allowed for the f-fluid, during
which the matter of the fluid propagates without interactions. 
The formation time 
is  associated with a finite time of string formation. 
Each of these fluids (the f-fluid after its formation) 
is governed by conventional hydrodynamic equations. The continuity equations 
for the baryon charge read 
   \begin{eqnarray}
   \label{eq8}
   \partial_{\mu} J_{\alpha}^{\mu} (x) &=& 0,
   \end{eqnarray}
for $\alpha=$p and t, where
$J_{\alpha}^{\mu}=n_{\alpha}u_{\alpha}^{\mu}$ is the baryon
current defined in terms of proper (i.e. in the local rest frame) net-baryon density $n_{\alpha}$ and
 hydrodynamic 4-velocity $u_{\alpha}^{\mu}$ normalized as
$u_{\alpha\mu}u_{\alpha}^{\mu}=1$. Eq.~(\ref{eq8}) implies that
there is no baryon-charge exchange between p-, t- and f-fluids, as
well as that the baryon current of the fireball fluid is
identically zero, $J_{\scr f}^{\mu}=0$. 
In fact, the latter is a forced assumption of the model in order to 
avoid introduction of additional phenomenological (and hence unknown) 
diffusion coefficients of the baryon charge between fluids. However, this 
assumption is not restrictive because a finite net-baryon density is 
provided by baryon-rich fluids which always overlap with the f-fluid,
see Fig. \ref{fig1a}.

Equations of the energy--momentum exchange between fluids are formulated
in terms of 
energy--momentum tensors $T^{\mu\nu}_\alpha$ of the 
fluids
   \begin{eqnarray}
   \partial_{\mu} T^{\mu\nu}_{\scr p} (x) &=&
-F_{\scr p}^\nu (x) + F_{\scr{fp}}^\nu (x),
   \label{eq8p}
\\
   \partial_{\mu} T^{\mu\nu}_{\scr t} (x) &=&
-F_{\scr t}^\nu (x) + F_{\scr{ft}}^\nu (x),
   \label{eq8t}
\\
   \partial_{\mu} T^{\mu\nu}_{\scr f} (x) &=&
- F_{\scr{fp}}^\nu (x) - F_{\scr{ft}}^\nu (x)
\cr
&+&
\int d^4 x' \delta^4 \left(\vphantom{I^I_I} x - x' - U_F
(x')\tau_f\right)
\cr
&\times&
 \left[F_{\scr p}^\nu (x') + F_{\scr t}^\nu (x')\right],
   \label{eq8f}
   \end{eqnarray}
{
where the energy--momentum tensor of the $\alpha$ fluid is defined in the perfect-fluid form 
$$
T^{\mu\nu}_\alpha = (\varepsilon_{\alpha}+P_{\alpha}) u_{\alpha}^{\mu} u_{\alpha}^{\nu} 
- g^{\mu\nu} P_{\alpha} 
$$
with $\varepsilon_{\alpha}$ and $P_{\alpha}$ 
being the proper energy density and the pressure of the $\alpha$ fluid, respectively, 
and 
}
the $F^\nu_\alpha$ are friction forces originating from
inter-fluid interactions. $F_{\scr p}^\nu$ and $F_{\scr t}^\nu$ in
Eqs.~(\ref{eq8p})--(\ref{eq8t}) describe energy--momentum loss of the 
baryon-rich fluids due to their mutual friction. A part of this
loss $|F_{\scr p}^\nu - F_{\scr t}^\nu|$ is transformed into
thermal excitation of these fluids, while another part $(F_{\scr
p}^\nu + F_{\scr t}^\nu)$ gives rise to particle production into
the fireball fluid (see Eq.~(\ref{eq8f})). $F_{\scr{fp}}^\nu$ and
$F_{\scr{ft}}^\nu$ are associated with friction of the fireball
fluid with the p- and t-fluids, respectively. 
Here $\tau_f$ is the formation time, and
   \begin{eqnarray}
   \label{eq14}
U^\nu_F (x')=
\frac{u_{\scr p}^{\nu}(x')+u_{\scr t}^{\nu}(x')}%
{|u_{\scr p}(x')+u_{\scr t}(x')|}
   \end{eqnarray}
is a  4-velocity the free-propagation of the produced fireball 
matter.
In fact, this is a velocity of the
fireball matter at the moment of its production. 
Accordingly to Eq.~(\ref{eq8f}), 
this matter 
gets formed only 
after the time span $U_F^0\tau_f$ upon the
production, and in 
different space point ${\bf x}' - {\bf U}_F (x') \ \tau_f$, as
compared to the production point ${\bf x}'$.

Making sums of Eq.~(\ref{eq8}) for different fluids ($\alpha =$ p, t and f) 
we arrive at the local  baryon charge conservation 
   \begin{eqnarray}
   \label{eq8a}
   \partial_{\mu} [J_{\scr p}^{\mu} (x)+J_{\scr t}^{\mu} (x)] = 0 
   \end{eqnarray}
 {with due account that $J_{\scr f}^{\mu}=0$ by the model assumption.}
Summation of Eqs.  (\ref{eq8p})-(\ref{eq8f}) gives us 
   \begin{eqnarray}
   \partial_{\mu} [T^{\mu\nu}_{\scr p} (x) + T^{\mu\nu}_{\scr t} (x) + T^{\mu\nu}_{\scr f} (x)] =
-\left[F_{\scr p}^\nu (x) + F_{\scr t}^\nu (x)\right]&&
\cr
+
\int d^4 x' \delta^4 \left(\vphantom{I^I_I} x - x' - U_F
(x')\tau_f\right)
 \left[F_{\scr p}^\nu (x') + F_{\scr t}^\nu (x')\right].&&
   \label{eq8b}
   \end{eqnarray}
The l.h.s. of this equation is nonzero, in general. It means that there is 
no energy-momentum conservation. This occurs because a part of the energy-momentum 
is stored in the still unformed f-fluid that does not take part in the 
hydrodynamic evolution. At later stages of the collision the loss term on the r.h.s. 
$\left[F_{\scr p}^\nu (x) + F_{\scr t}^\nu (x)\right]$ 
becomes zero because the baryon-rich (p and t fluids) are either mutually stopped and hence unified
or they are spatially separated. When in addition the formation of the f-fluid is completed, 
the gain term on the r.h.s., i.e. the integral term, also becomes zero. Then  
the r.h.s.  of this equation 
becomes zero. Thus, we arrive at the local  energy-momentum conservation.

Standard manipulations \cite{Land-Lif,Rischke:1998fq} with Euler equations (\ref{eq8p})-(\ref{eq8f}) 
give us entropy production in each fluid 
   \begin{eqnarray}
   \partial_{\mu} (u^{\mu}_{\scr p} s_{\scr p}) &=&
u_{{\scr p}\mu}\,(-F_{\scr p}^\mu + F_{\scr{fp}}^\mu),
   \label{eq8ps}
\\
   \partial_{\mu} (u^{\mu}_{\scr t} s_{\scr t}) &=&
u_{{\scr t}\mu}\,(-F_{\scr t}^\mu + F_{\scr{ft}}^\mu),
   \label{eq8ts}
\\
   \partial_{\mu} (u^{\mu}_{\scr f} s_{\scr f}) &=&
u_{{\scr f}\mu}\,\Bigl\{ - F_{\scr{fp}}^\mu - F_{\scr{ft}}^\mu 
\cr
&+&
\int d^4 x' \delta^4 \left(\vphantom{I^I_I} x - x' - U_F
(x')\tau_f\right)
\cr
&\times&
 \left[F_{\scr p}^\mu (x') + F_{\scr t}^\mu (x')\right]
\Bigr\},
   \label{eq8fs}
   \end{eqnarray}
where $s_{\alpha}$ is a proper entropy density of the $\alpha$ fluid. 
Though, it is not necessarily to use these equations to calculate the entropy production. 
It is easier to use the thermodynamic relation 
   \begin{eqnarray}
s_{\alpha} = \frac{1}{T_{\alpha}} (\varepsilon_{\alpha} + P_{\alpha} - n_{\alpha} \mu_{\alpha})
   \label{sa}
   \end{eqnarray}
in order to determine $s_{\alpha}$. Here, $T_{\alpha}$ and $\mu_{\alpha}$
are the temperature and the baryon chemical 
potential of the $\alpha$ fluid, respectively.
{
Note that $n_{\scr f}= \mu_{\scr f}=0$.
}
All these quantities are known from solution of the 3FD equations  (\ref{eq8})-(\ref{eq8f}). 
The total entropy  is then calculated by integration over volume of the system 
   \begin{eqnarray}
S = \sum_{\alpha} \int dV  u^0_{\alpha}  s_{\alpha} 
   \label{stot}
   \end{eqnarray}

However, this is not quite consistent definition of the entropy because the entropy
is not always an extensive  quantity. 
The entropy is indeed an extensive  quantity with respect to matter contained 
in several non-overlapping volumes or to different particle species in the same volume, 
though not with respect to identical particles in the same volume which are artificially 
subdivided into several subgroups. 
For instance, let us consider pions which are present in all three fluids. 
Let the distribution function of pions over momenta ($p$) in a small volume ($\delta V$) be 
   \begin{eqnarray}
   \label{f-dist}
f(p) = f_{\scr p}(p) + f_{\scr t}(p) + f_{\scr f}(p),
   \end{eqnarray}
i.e. the pions are artificially distributed between three fluids in the same volume. 
Assuming that these pions compose an ideal gas, their entropy should be calculated as follows 
   \begin{eqnarray}
   \label{ds-real}
\delta S = 
-\delta V \int d^3 p (f_{\scr p} + f_{\scr t} + f_{\scr f}) [\ln(f_{\scr p} + f_{\scr t} + f_{\scr f}) -1]         
\hphantom{nn}  
   \end{eqnarray}
which generally is not equal to 
   \begin{eqnarray}
   \label{ds-3fliud}
\delta S{\scr{subgroups}} = 
-\delta V \sum_{\alpha} 
 \int d^3 p  f_\alpha(\ln f_\alpha-1), 
   \end{eqnarray}
i.e. the sum of entropies of separate subgroups. The account of Bose statistics, 
i.e. additional terms of $(1-f_p)\ln(1-f_p)$, etc., does not change the situation. 
In particular, this fact is the origin of the so-called Gibbs paradox in thermodynamics. 
The maximal difference between $\delta S$ and $\delta S{\scr{subgroups}}$ is reached
when $f_{\scr p} = f_{\scr t} = f_{\scr f}$. 
Then $(\delta S - \delta S{\scr{subgroups}})/\delta N = - \ln 3$, where $\delta N$ is the total number of particles 
(i.e. pions, in this example) in volume $\delta V$.

In fact, the 3FD entropy defined by Eq. (\ref{stot}) is of the same nature as that in Eq. (\ref{ds-3fliud}).  
Therefore, it can overestimate a true entropy, at most, by term $\sim N$ where $N$ is the total number of hadrons 
produced in the collision. This systematic error should be kept in mind when inspecting results 
of the 3FD calculations of the entropy. 
In principle, this problem of the entropy definition could be remedied by directly using  
definition (\ref{ds-real}) for the entropy of the composed system. Moreover, this 
definition can be extended to the case of quasiparticles in a mean field. The latter  
is in fact the picture underling the used EoS's. However, this improvement would 
make the entropy computation enormously complicated. Therefore, we prefer a simpler 
approximate calculation complemented by an error estimate of such a simplification.

The physical input
of the present 3FD calculations is described in detail in
Ref.~\cite{Ivanov:2013wha}. The friction between fluids was fitted to reproduce
the stopping power observed in proton rapidity distributions for each EoS, 
as it is described in  Ref. \cite{Ivanov:2013wha} in detail.
The main difference concerning the 
f-fluid in considered alternative scenarios
consists in different formation times: $\tau$ = 2 fm/c for the
hadronic scenario and
$\tau$  = 0.17 fm/c for scenarios involving 
the deconfinement transition \cite{Ivanov:2013wha}.
Large formation time within the hadronic scenario was chosen in order 
to reproduce hadronic yields at SPS energies.
This was done in line with a principle of fair treatment of any EoS: 
any possible uncertainties in the parameters are treated in favor of the EoS,
i.e. for each EoS the dynamical parameters of the model are chosen (within their uncertainty range) 
in such a way that  
the best possible reproduction of observables is achieved with this EoS. 
Though, even this relatively large formation time did not allow us to 
reproduce these yields above the SPS energies within the hadronic scenario \cite{Ivanov:2013yqa}.

In order to illustrate the dynamics of the nuclear collision emerging from 
the above described model, Fig. \ref{fig1a} presents the 
time evolution of the proper (i.e. in the local rest frame) baryon density   
(upper row of panels), the proper energy density 
(middle row of panels) and the proper energy density of the f-fluid  
(lower row of panels) in the $xz$ plane (i.e. the reaction plane)
of central Au+Au collision within the crossover
scenario at $\sqrt{s_{NN}}=$ 19.6 GeV. 
The total baryon and energy densities were calculated performing an 
artificial unification of all the individual fluids. 
At the initial stage of the reaction 
the p- and t-fluids interpenetrate each other and 
produce the f-fluid due to mutual friction. Afterwards 
these fluids get either  spatially separated again or 
completely (at lower collision energies) or partially 
(at higher collision energies) unified.  At the same time
the f-fluid gets partially absorbed by the p- and t-fluids 
but keeps it identity till the very end  
of the collision (i.e. freeze-out).

Though the f-fluid is produced in the midrapidity region, 
its hydrodynamical (quasi-1D) expansion stretches the 
fluid along the collision axis ($z$). In addition, 
the p- and t-fluids carry it along with them due to 
friction interaction thus additionally stretching 
the f-fluid along the collision axis.
Thus, the f-fluid overlaps with the baryon-rich (p- and t-) fluids 
till the final stage, as it is seen from  Fig. \ref{fig1a}. 
Moreover, the f-fluid does not give a dominant contribution to the total 
energy density. 
Similar dynamical pattern takes place up to the highest collision 
energies considered in this paper.

\section{Effective viscosity}
\label{viscosity}

The main idea of estimating an effective shear viscosity in a nuclear collision consists 
in associating the entropy production within the 3FD simulation with the effect of 
the viscous dissipation within the standard (first-order) viscous hydrodynamics. 
We rely on the first-order viscous hydrodynamics, rather then on modern versions of 
the second-order viscous formulations, because this is quite enough for our purpose, 
i.e. a rough estimate of the effective viscosity.

In fact, the 3FD dissipation is directly related neither to the shear viscosity nor 
other transport coefficients, i.e. the bulk viscosity and thermal conductivity. 
The dissipation due to these transport coefficients takes place only when gradients
of the collective velocity, temperature and chemical potential exist \cite{Land-Lif,Rischke:1998fq}. 
The 3FD dissipation can, in principal, occur even without any gradients,  
e.g. in two homogeneous counter-streaming media. 
Though the real evolution 
of the nuclear collision gives rise to such gradients. Thus, we can express the 
3FD dissipation in familiar terms by associating it with the shear viscosity. 
The shear viscosity is chosen among other transport coefficients only because 
the dissipation in heavy-ion collisions is traditionally discussed in terms of this quantity.

Let us briefly recollect the  Landau-Lifshitz formulation of the standard dissipative 
hydrodynamics \cite{Land-Lif,Rischke:1998fq}, 
where hydrodynamic velocity is associated with the energy flow.   
Dissipative terms modify the baryon current $J_{\scr B}^{\mu}$ and the 
energy--momentum tensor $T^{\mu\nu}$ as follows
   \begin{eqnarray}
   \label{J-dissipative}
J_{\scr B}^{\mu} &=& u^{\mu} n_{\scr B} + q^{\mu} \quad \quad \quad 
\\
   \label{T-dissipative}
T^{\mu\nu} &=& 
(\varepsilon + P) \ u^{\mu} \ u^{\nu} -g^{\mu\nu} P + \pi^{\mu\nu}, 
   \end{eqnarray}
where
the heat flow, $q^{\mu}$, and the stress tensor, $\pi^{\mu\nu}$,
   \begin{eqnarray}
   \label{pi-dissipative}
\pi^{\mu\nu} &=& \eta 
\left(
\partial^{\mu}u^{\nu} + \partial^{\nu}u^{\mu}
-u^{\mu} u_{\lambda}\partial^{\lambda}u^{\nu} 
-u^{\nu} u_{\lambda}\partial^{\lambda}u^{\mu} 
\right)
\cr
&+&
\left(\zeta - \frac{2}{3} \eta \right)
\left(g^{\mu\nu} -u^{\mu} u^{\nu} \right)\partial_{\lambda}u^{\lambda}
\\
   \label{nu-dissipative}
q^{\mu} &=& \kappa \left(\frac{nT}{\varepsilon + P}\right)^2
\left(\partial^{\mu} - u^{\mu} \ u^{\nu} \partial_{\nu}\right)
\left(\frac{\mu}{T}\right)
   \end{eqnarray}
are expressed in terms of a shear viscosity, $\eta$, bulk viscosity, $\zeta$, 
and thermal conductivity, $\kappa$.  Here $\mu$ and $T$ are the baryon chemical 
potential and temperature, respectively. 
Then the equations of motion (the continuity and Navier–Stokes equations) 
result in the equation for the entropy production
   \begin{eqnarray}
   \label{s-dissipative}
\partial_{\mu} s^{\mu} = - q_{\mu} \partial^{\mu} \left(\frac{\mu}{T}\right) 
+ 
\frac{1}{T}\pi_{\mu\nu} \partial^{\mu} u^{\nu} 
   \end{eqnarray}
with nonnegative r.h.s.,  that ensures the second law of thermodynamics.

If we know the entropy production rate (cf. previous section), 
from Eq. (\ref{s-dissipative}) we can determine only a single transport coefficient. 
Let it be the shear viscosity, $\eta$, because the dissipation in heavy-ion 
collisions is traditionally discussed precisely in terms of this quantity. 
Then we have to put all other coefficients to be zero, 
$\zeta=\kappa=0$. If $\kappa=0$, the heat flow also vanishes, $q^{\mu}=0$, and hence 
hydrodynamic velocity $u^{\mu}$ is associated with baryon flow, see Eq. (\ref{J-dissipative}). 
Had we started from the Eckart formulation of the dissipative 
hydrodynamics \cite{Rischke:1998fq}, where hydrodynamic velocity is 
associated with baryon flow from the beginning, 
we would have no other choice all the more so.

Let us start to construct a phenomenological correspondence between the 3FD model and the 
conventional dissipative hydrodynamics. 
As it is dictated by the conventional dissipative hydrodynamics with only
shear viscosity,
we define 
the hydrodynamic velocity $u^{\mu}$ as follows 
   \begin{eqnarray}
   \label{u-dissipative}
n_B u^{\mu} = n_p u_p^{\mu} + n_t u_t^{\mu}
   \end{eqnarray}
in terms of 3FD net-baryon densities and velocities, i.e. as if the 
3FD baryon-rich fluids are unified.  The f-fluid does not take part 
in this definition. 
The mean temperature, $T$, that is also required by Eq. (\ref{s-dissipative}), 
is defined proceeding from common sense, i.e. it is defined
as a local  proper-energy-density-weighted temperature
   \begin{eqnarray}
   \label{Tm-dissipative}
T = \sum_\alpha T_{\alpha} \varepsilon_{\alpha} \Big/ \sum_\alpha \varepsilon_{\alpha} 
   \end{eqnarray}
where $\varepsilon_{\alpha}$ is a proper energy density of the $\alpha$ fluid. 

Integrating Eq. (\ref{s-dissipative}) over volume, $V$, and keeping in mind 
that only $\eta$ in nonzero, we arrive at 
   \begin{eqnarray} 
   \label{S(t)}
\frac{dS}{dt}  &=&  V \left\langle \frac{1}{T}\pi_{\mu\nu} \partial^{\mu} u^{\nu} \right\rangle 
\cr
&\approx& 
\left( \frac{V\langle\eta\rangle}{S}\right) \frac{S}{\langle T\rangle}
\left\langle \frac{1}{\eta}\pi_{\mu\nu} \partial^{\mu} u^{\nu} \right\rangle,  
   \end{eqnarray}
where $\langle ...\rangle$ denotes average over the volume. 
Taking into account that $s^{\mu} = s u^{\mu}$, where $s$ is the proper entropy density, 
and hence $S = V \langle s u^{0}\rangle \approx  V \langle s\rangle  \langle u^{0}\rangle$, 
we obtain    
   \begin{eqnarray} 
   \label{eta-s}
\frac{\langle\eta\rangle}{\langle s\rangle}
\approx
\left(\frac{1}{S}\frac{dS}{dt}\right)   
\langle T\rangle\langle u^{0}\rangle
\Big/
\left(
\left\langle \frac{1}{\eta}\pi_{\mu\nu}\right\rangle \left\langle\partial^{\mu} u^{\nu} \right\rangle   
\right).
   \end{eqnarray}
The last step, i.e. $
\left\langle \pi_{\mu\nu} \partial^{\mu} u^{\nu} \right\rangle
\approx
\left\langle \pi_{\mu\nu}\right\rangle \left\langle\partial^{\mu} u^{\nu} \right\rangle$, 
has been done in order to facilitate numerical evaluation of terms with time derivatives. 
This expression, together with definitions (\ref{u-dissipative}) and (\ref{Tm-dissipative}), 
is the final relation that is used for estimation of the $\eta/s$ ratio.

\section{Results}
\label{Results}

\begin{figure}[tbh]
\includegraphics[width=8.5cm]{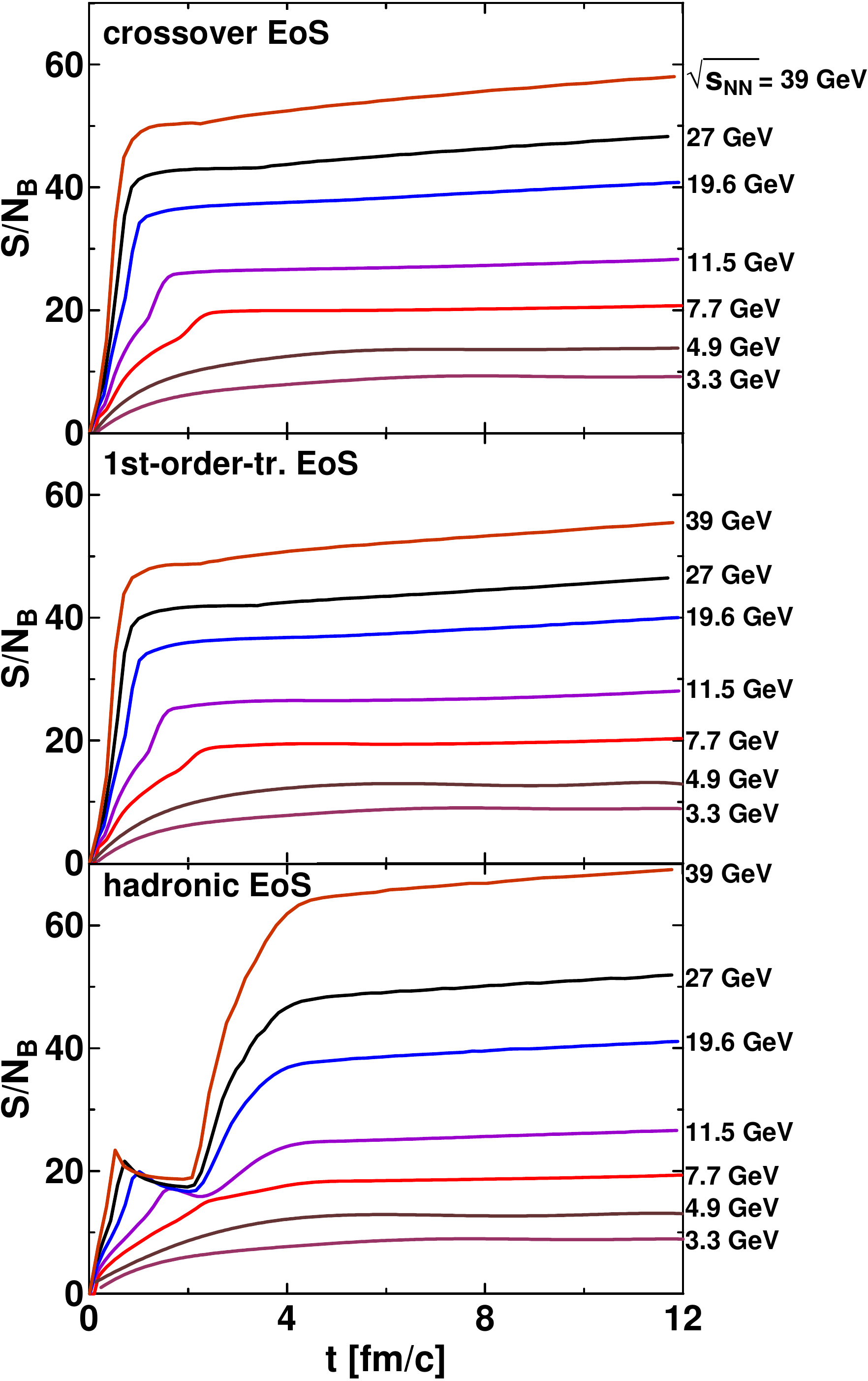}
 \caption{
Specific entropy per net baryon ($N_B=2A=$ 394) generated in central Au+Au collisions  at various energies $\sqrt{s_{NN}}=$ 3.3--39 GeV
within different scenarios of the 3FD simulations: a purely hadronic EoS \cite{gasEOS}  
and two versions of the EoS involving the   deconfinement
 transition \cite{Toneev06}---a first-order phase transition  
and a smooth crossover one.
}
\label{fig1}
\end{figure}

The 3FD simulations of central Au+Au collisions at energies 
3.3 GeV  $\le \sqrt{s_{NN}}\le$ 39 GeV were performed without freeze-out. 
The freeze-out in the 3FD model removes the frozen out matter from the hydrodynamical 
evolution \cite{Russkikh:2006aa,Ivanov:2008zi}, hence the entropy growth of the hydrodynamically 
evolving matter is not guarantied. 
Therefore, in order to keep all the matter in the 
consideration the freeze-out was turned off.
Results of the calculations of the entropy production are demonstrated in Fig. \ref{fig1}.
The time is counted in the c.m. frame of colliding nuclei. 
As seen, the entropy rapidly grows at the early stage of the collision. 
This growth is somewhat 
irregular in the deconfinement-transition scenarios and even  changes into temporal fall in the
hadronic scenario. These are effects of the formation time of the f-fluid. The unformed part of the 
matter drops out of the fluid evolution for the period of its formation, i.e. the formation time. 
Therefore, the fluid system becomes an open one which exchanges the energy and momentum with an 
external bath of the unformed matter. If the formation time is long, like in the hadronic scenario 
($\tau_f=$ 2 fm), the fluid system is strongly affected by the exchange with this external bath.

At the initial stage of the reaction all three fluids coexist in the same 
space-time region, thus describing a certain {\em nonequilibrium} state 
of the matter. As seen from Fig. \ref{fig1}, this is the stage of the 
fastest entropy growth that takes time comparable with that required by two 
nuclei to pass each other moving with their initial velocities, of course, 
with due account of the Lorentz contraction of the nuclei in the c.m. frame.  
Due to this Lorentz contraction this initial stage becomes shorter with 
the collision energy rise. 

This fast initial entropy growth is followed by a longer stage of quite 
moderate entropy generation. 
At this stage the p- and t-fluids are either spatially separated or unified, 
while the f-fluid still overlaps 
with the baryon-rich (p- and t-) fluids, 
 {see Fig. \ref{fig1a}}
%
Therefore, the friction between 
the f-fluid and the baryon-rich fluids still courses the dissipation and 
hence the entropy growth.

The numerics of the model also contributes to the entropy production. 
In Refs. \cite{Horvat:2010dk,Csernai:2011qq} this question was studied 
%
for the numerical scheme based on the Particle-in-Cell (PIC) method. 
The same scheme is used in the 3FD simulations. 
The authors of Refs. \cite{Horvat:2010dk,Csernai:2011qq} 
used the Eulerian cells from $dx=dy=dz=$ 0.585 fm to 0.35 fm
and concluded that 
the numerical viscosity of their calculations
is small, i.e. 0.1 in terms of the $\eta/s$ ratio, based on the small Eulerian cells.
Moreover, the numerical viscosity decreases with the Eulerian cell decrease. 
In our simulations we use even smaller cells: from $dx=dy=dz=$ 0.192 fm at 
$\sqrt{s_{NN}}=$ 3.3 GeV to $dx=dy=dz=$ 0.0261 at $\sqrt{s_{NN}}=$ 39 GeV. 
This implies that the numerical viscosity at the expansion stage of the collision is smaller 
than 0.1 in terms of the $\eta/s$ ratio in the 3FD calculations. 
At the compression stage of the collision we have approximately the same number of cells per 
the longitudinal size of the system (not less than $\sim$ 40)
at all considered collision energies. Therefore, at 
this stage we expect the numerical viscosity (i.e. the $\eta/s$ ratio) 
of the order of 0.1, i.e. similar to that in 
Refs. \cite{Horvat:2010dk,Csernai:2011qq}.

\begin{figure}[tbh]
\includegraphics[width=7.5cm]{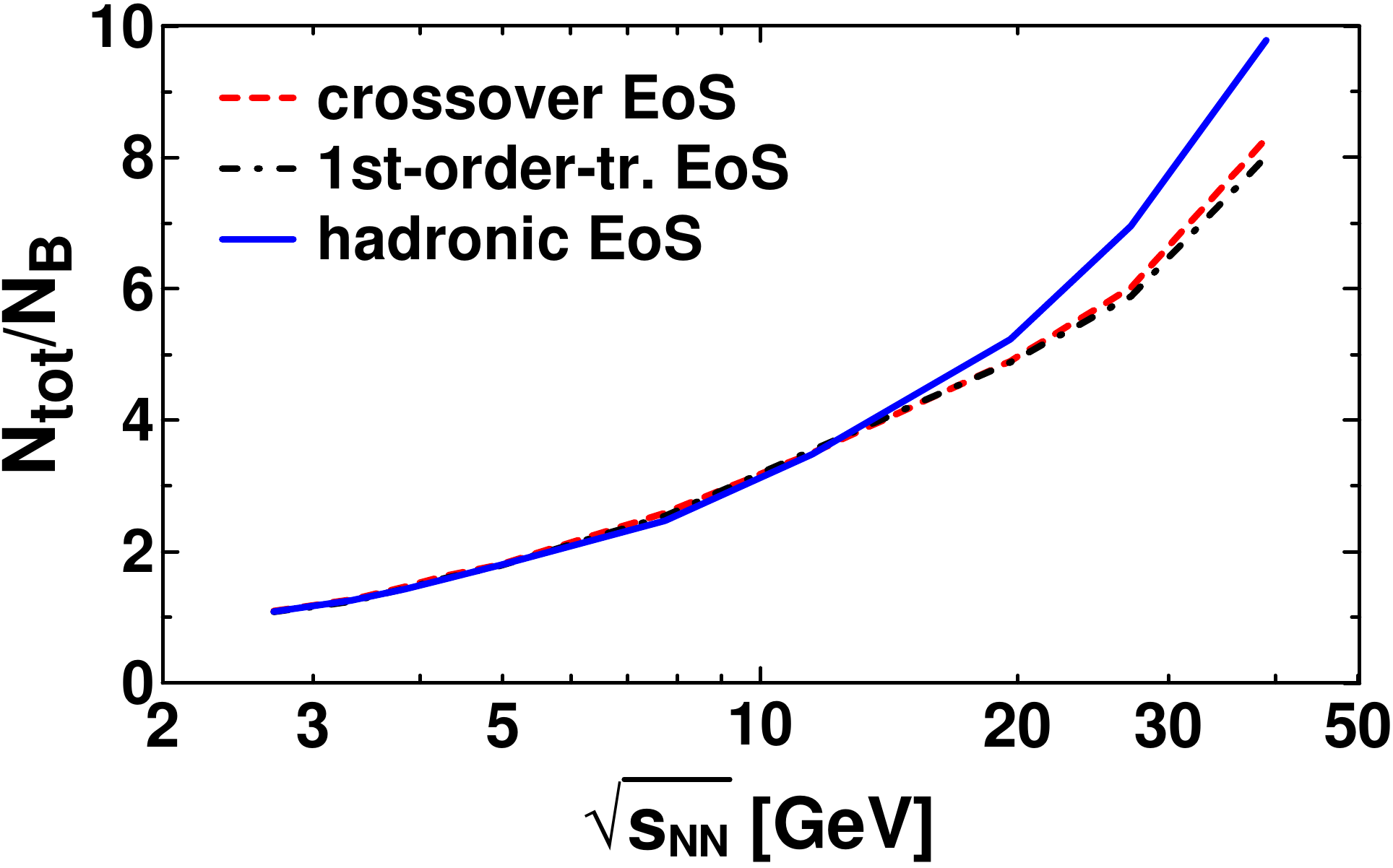}
 \caption{
Number of  hadrons per net baryon ($N_B=2A=$ 394) produced in central Au+Au collisions  
as a function of collision energy 
within different scenarios of the 3FD simulations. 
}
\label{fig2}
\end{figure}

Another source of uncertainty is the entropy definition in the 3FD model, cf. Eq. (\ref{stot}). 
As it was mentioned in  Sect. \ref{Model}, this systematic error can be 
estimated as $\delta S\sim N$ where $N$ is the total number of hadrons 
produced in the collision. In Fig. \ref{fig2} the number of produced hadrons, 
including resonances, is presented as a function of the collision energy.  
To be precise, this is the number of hadrons before their strong decays, 
because precisely this quantity matters for the thermodynamics. 
After the strong decays the number of produced hadrons becomes larger 
because of two- and three-body decays. Comparing Figs. \ref{fig1} and \ref{fig2}, 
we see that the error of the entropy definition at the final stage of the evolution 
amounts $\sim$10\% at all collision energies and scenarios. 

As seen from Fig. \ref{fig1}, the entropy produced to the end of the collision is 
approximately the same in all considered scenarios. Only in the hadronic scenario 
at $\sqrt{s_{NN}}>$ 20 GeV
the entropy exceeds the values reached 
in the confinement scenarios. The latter results in an overestimation of abundances of 
secondary particles (in particular, pions)     
within the  hadronic scenario \cite{Ivanov:2013yqa} which  
is clearly seen already from 
Fig. \ref{fig2}. The entropy production in our simulations is in good agreement with 
similar results obtained in the early (Frankfurt) version of the three-fluid dynamics
\cite{Brac98}. In Ref. \cite{Brac98} calculations were performed at $\sqrt{s_{NN}}<$ 20 GeV. 

%
\begin{figure}[tbh]
\includegraphics[width=8.5cm]{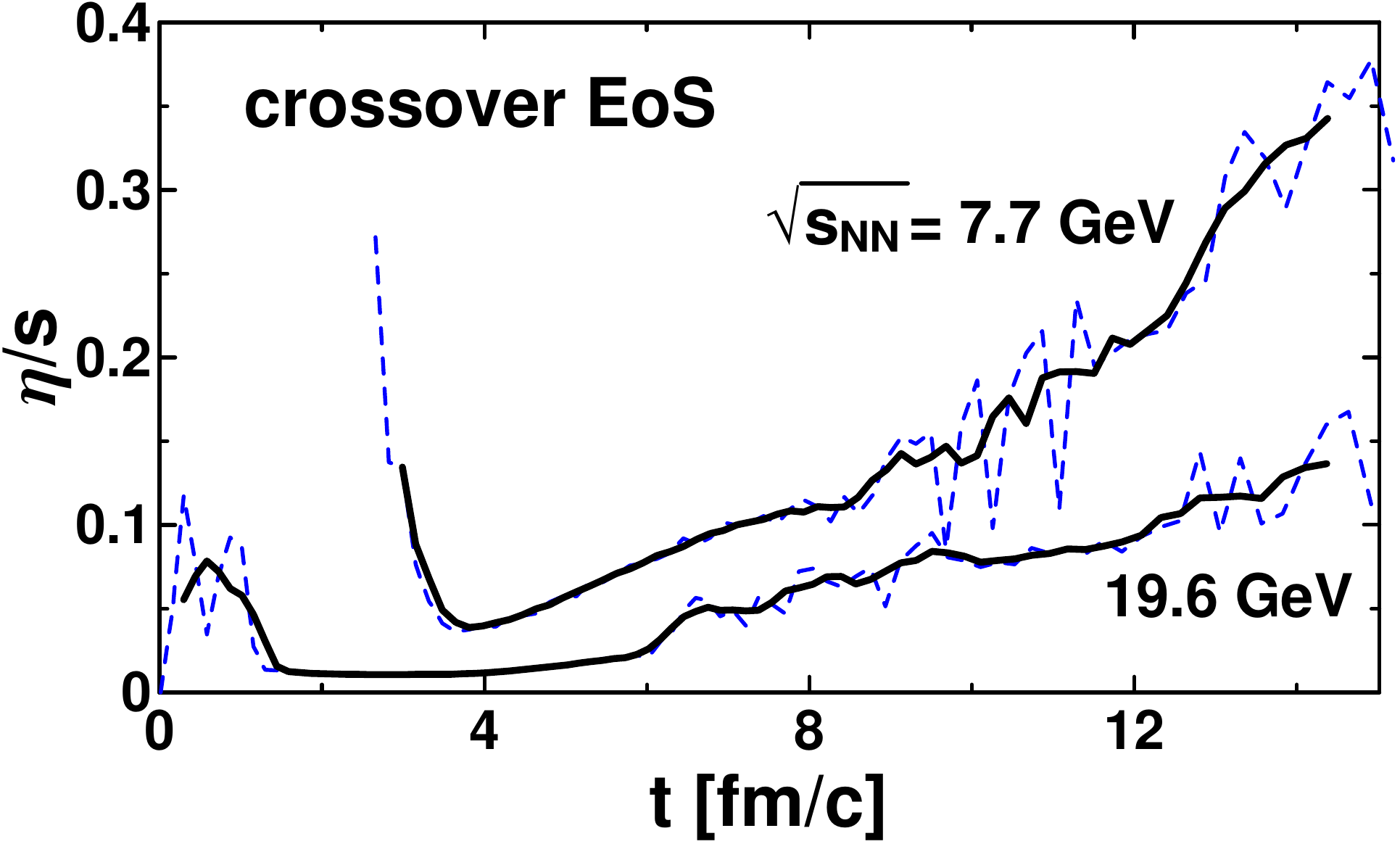}
 \caption{
The $\eta/s$ ratio as a function of time for central Au+Au collisions at 
two collision energies $\sqrt{s_{NN}}=$ 7.7 and 19.6 GeV 
within the crossover scenario.  Results of direct calculations based on 
Eq. (\ref{eta-s}) are displayed by thin dashed lines, 
the smoothed results of running average over five time steps, by bold solid lines. 
}
\label{fig3}
\end{figure}

The $\eta/s$ ratio evaluated on the basis of Eq. (\ref{eta-s}) is a function of time 
because all the quantities entering Eq. (\ref{eta-s}) depend on the time evolution 
of the colliding system. 
An example of results on estimated $\eta/s$ ratio is presented in Fig. \ref{fig3}.
As seen from Fig. \ref{fig3}, $\eta/s$ results strongly fluctuate in time (thin dashed 
curves in Fig. \ref{fig3}). 
This is a consequence the numerical calculation of the derivatives that aways courses 
a loss of accuracy. The amplitude of these fluctuations characterizes the accuracy of 
the $\eta/s$ estimate. In view of this poor accuracy and a very approximate nature of 
Eq. (\ref{eta-s}) itself, the present results on the $\eta/s$ ratio should be considered as an 
order-of-magnitude estimation. For the sake of the graphic representation, we apply running 
average procedure to the results of the direct calculation in such a way that 
the $\eta/s$ ratio is averaged over each sequential five time steps. Though these running average results 
(bold solid curves in Fig. \ref{fig3}) are not completely smooth, they exhibit much 
weaker fluctuations.

%
\begin{figure}[bht]
\includegraphics[width=8.5cm]{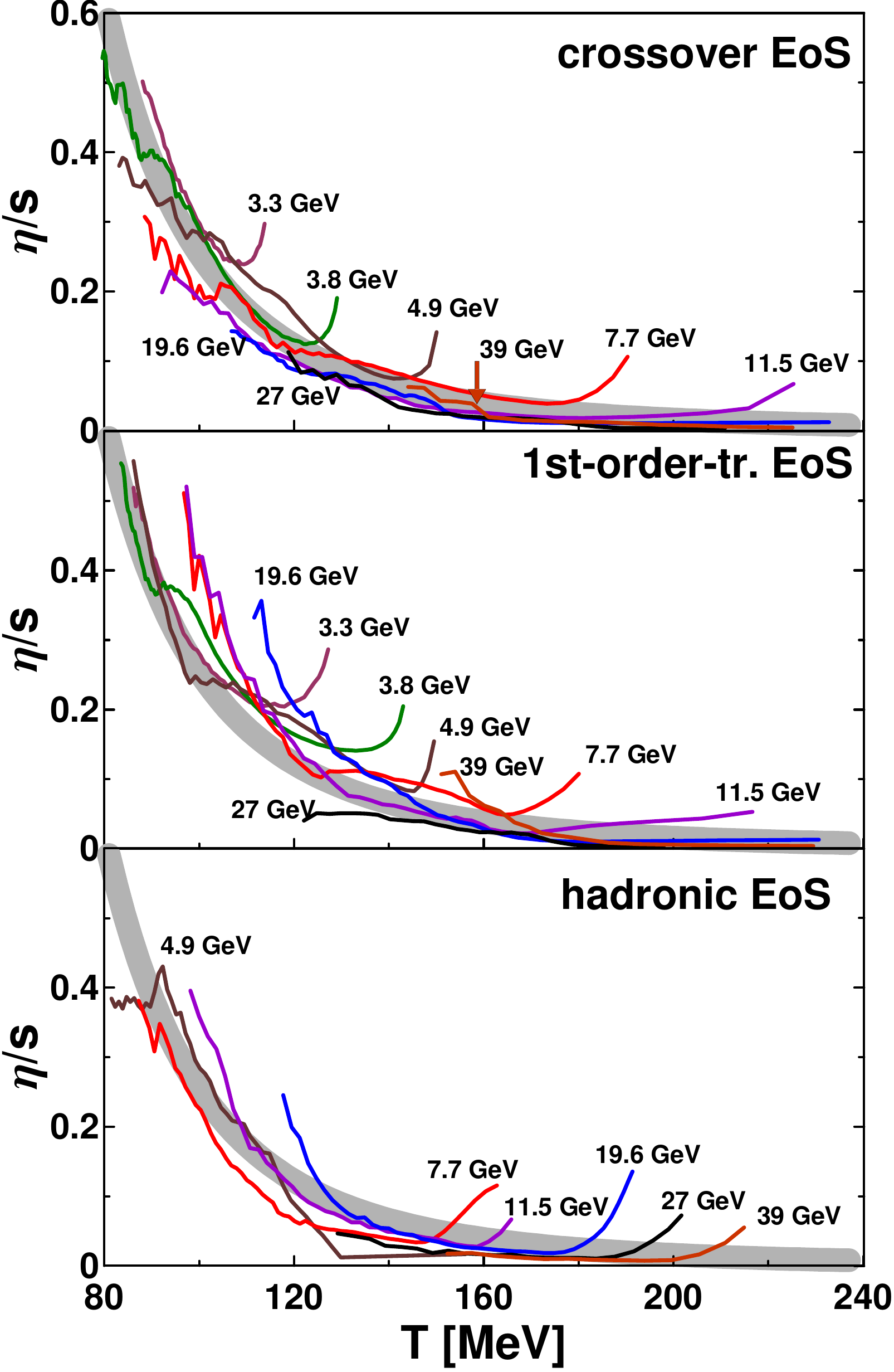}
 \caption{
The smoothed $\eta/s$ ratio as a function of temperature along trajectories of central Au+Au collisions at 
various collision energies $\sqrt{s_{NN}}$ 
within different scenarios. 
The gray band in all the panel is the function $(T_0/T)^4$, where $T_0=$ 71 MeV. 
}
\label{fig4}
\end{figure}
%
\begin{figure}[hbt]
\includegraphics[width=8.5cm]{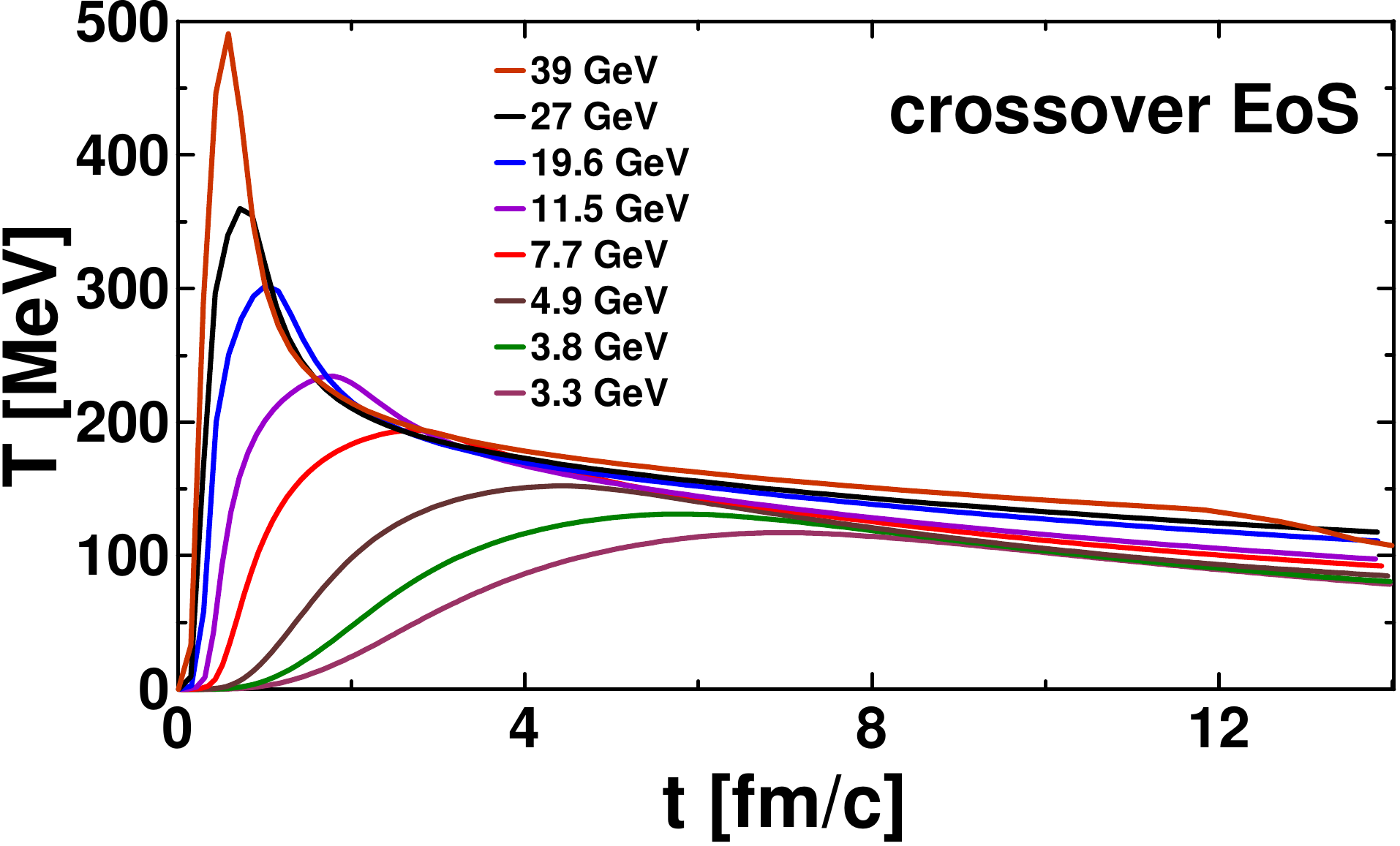}
 \caption{
The mean temperature of the system as a function of time for central Au+Au collisions at 
various collision energies $\sqrt{s_{NN}}$ 
within crossover scenario.}
\label{fig5}
\end{figure}

In Fig. \ref{fig4} the smoothed results on the $\eta/s$ ratio are presented   
as a functions of the mean temperature $\langle T\rangle$ 
[cf. Eqs.  (\ref{Tm-dissipative})-(\ref{eta-s})] along trajectories of central Au+Au collisions at 
various collision energies. 
These are based on simulations  
within three different scenarios. 
Only expansion stage of the collisions is displayed in this figure, 
because the system is close to equilibrium at this stage. 
At the initial strongly nonequilibrium stage the description in terms  
of the viscosity is inapplicable. 
In particular, the early-stage wiggle in the entropy production  
resulting from formation-time delay is beyond the range of Fig. \ref{fig4}. 
The results that manifest fluctuations comparable with  
the scale of the plot are omitted. Large fluctuations take place at the very late 
expansion stages, when the system becomes very dilute.

Another source of already systematic error is the entropy definition in the 3FD model, cf. Eq. (\ref{stot}). 
As it was shown above, this entropy error   amounts $\sim$10\% for all collision
energies and scenarios. In sect. \ref{Model} it was demonstrated that $S_{\rm 3FD} \simeq S_{\rm true} + N$, 
where $S_{\rm true}$ is a true entropy, $S_{\rm 3FD}$ is the entropy calculated by means of Eq. (\ref{stot}) 
and $N$ is the total particle number produced in the collision, note that $N \propto S_{\rm true}$. 
Therefore, we can rewrite entropy as $S_{\rm 3FD} \simeq (1+\delta)S_{\rm true}$ where $\delta \sim 0.1$. 
In particular, it means that $dS/dt$, entering the definition of the effective $\eta/s$ ratio of Eq. (\ref{eta-s}), 
also has the accuracy of $\sim$10\%. If we further take into account that $\eta/s$ is determined by 
combination $(1/S_{\rm true})dS_{\rm true}/dt$, we readily see that the 
$\delta$ error cancels out in Eq. (\ref{eta-s}). However, to be on the safe side, we assume that the 
systematic error  of the $\eta/s$ ratio due to the 3FD entropy definition is $\sim$10\%.

We plot the $\eta/s$ ratio as a function of the mean temperature rather than time 
because the temperature is a natural argument of the $\eta/s$ quantity. 
The mean temperature is also a function of time. 
The time dependence of the mean temperature is illustrated in Fig.  \ref{fig5} 
for the crossover scenario. 
The temperature has a physical meaning only for a sufficiently thermalized system, 
i.e.  at the stage when the fast entropy growth is over, see Fig. \ref{fig1}, or 
to the left of the $\eta/s$ minimum, see Fig. \ref{fig4}, in terms of 
the $\eta/s$ ratio. This stage of the gradual decrease of the temperature, see Fig. \ref{fig5}, 
is very similar for all considered scenarios.

At high temperatures ($T>$ 160 MeV) in collisions at $\sqrt{s_{NN}}>$ 10 GeV, the  $\eta/s$ ratio happens to 
be noticeably smaller that the conjectured  lowest bound for this quantity $1/(4\pi)$ 
\cite{Kovtun:2004de}, i.e. so called KSS bound. 
These small values $\eta/s$ should be considered as  
a property of the 3FD model, even though the present $\eta/s$ estimate is very rough. 
At final stages of the expansion%
\footnote{We avoid calling these stages 
the freeze-out ones because the freeze-out in the 3FD model is a continues process that 
takes place during the whole expansion stage \cite{Russkikh:2006aa,Ivanov:2008zi}.}
the  $\eta/s$ ratio possesses quite reasonable values---from $\sim$0.05 at highest 
considered energies to 0.5 at lowest ones. This range approximately agrees with 
those reported in Ref. \cite{Khvorostukhin:2010aj}
for the freeze-out in the statistical model.

The viscosity is meaningful when nonequilibrium is weak. 
Therefore, it should be analyzed at the expansion stage of the collision following 
the stage of the fast entropy growth, see Fig. \ref{fig1}. 
In terms of the $\eta/s$ ratio of Fig. \ref{fig4}, 
the expansion stage takes place at lower temperatures up to the minimum of 
the $\eta/s$ ratio.  
The $\eta/s$ curves are continued to higher temperatures after the minimum 
only for the sake of convenience of their labeling. 
As seen from Fig. \ref{fig4}, the temperature dependence of the $\eta/s$ ratio 
at the expansion stages of collisions at various collision energies
is very similar within different scenarios. This dependence is approximately described 
by $1/T^4$ low, i.e. this ratio decreases with the temperature rise, as it is commonly expected. 
It is important to emphasize that this is the $T$-dependence along dynamical trajectories 
of collisions, along which the mean proper net-baryon density, $n_B$, also changes.

%
\begin{figure}[tbh]
\includegraphics[width=8.5cm]{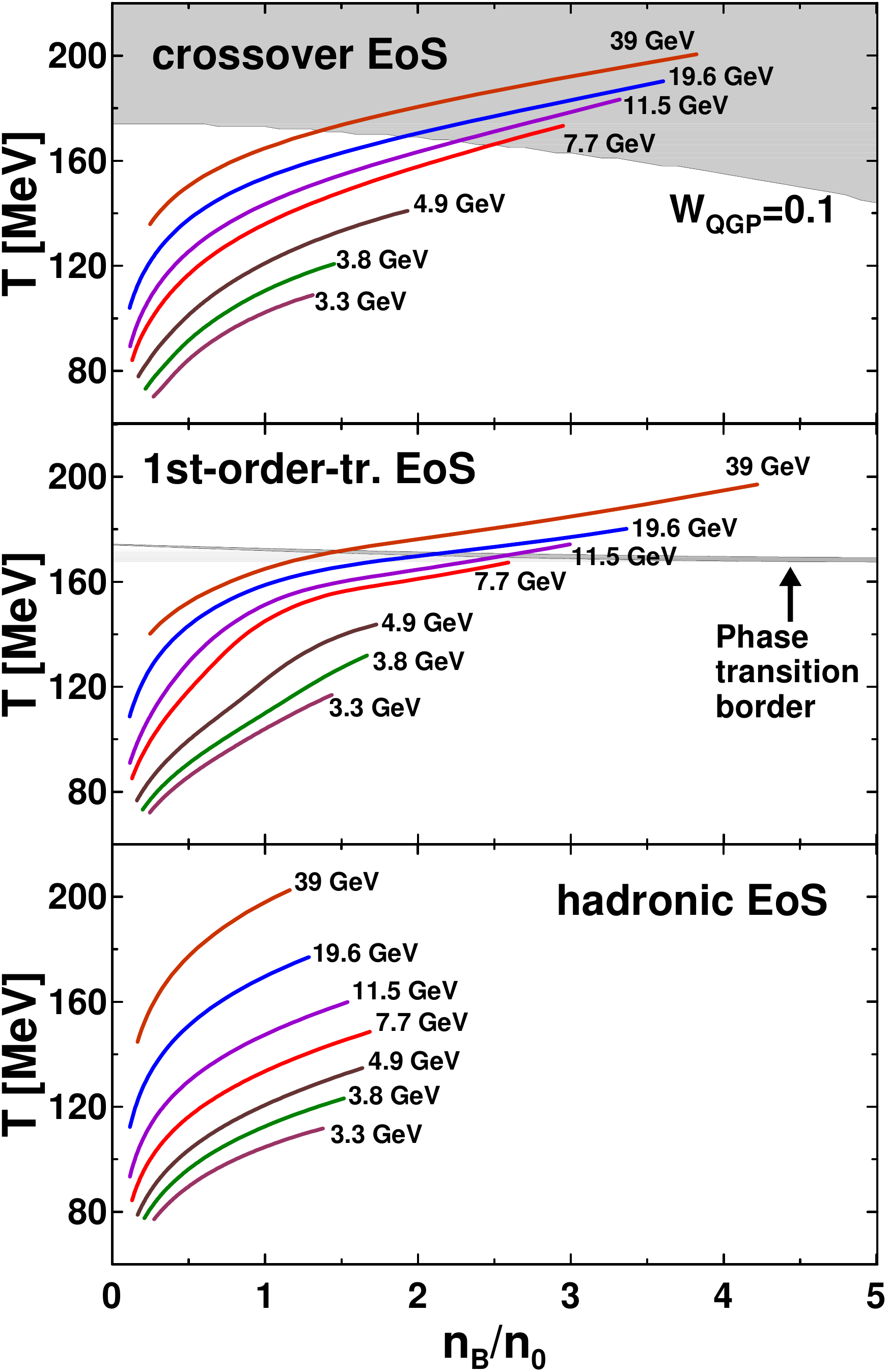}
 \caption{
Dynamical trajectories of expansion stages 
of central Au+Au collisions at various collision energies $\sqrt{s_{NN}}$ 
and within different scenarios in terms of the mean proper net-baryon density, $n_B$, 
(in units of the normal nuclear density, $n_0$) 
and the mean temperature averaged over the whole system of colliding nuclei. 
The shaded area in the crossover-EoS panel is the region where the QGP fraction, 
$W_{\rm{QGP}}$, exceeds the value of 0.1. For the first-order-transition EoS, 
the mixed phase region is displayed by the shaded area. 
}
\label{fig6}
\end{figure}

The dynamical trajectories of expansion stages 
of central Au+Au collisions at various collision energies  
and within different scenarios are displayed in Fig. \ref{fig6}
in terms of the mean proper net-baryon density 
and the mean temperature which are averaged over the whole system. 
The collision evolution proceeds from the top-right ends of the 
trajectories to the left-bottom ends, i.e. from high densities 
and temperatures to low ones. 
For the deconfinement scenarios the borders of the deconfinement transition 
are displayed: the border above which the QGP fraction, 
$W_{\rm{QGP}}$, exceeds the value of 0.1 for the crossover EoS, and 
the mixed phase region for the first-order-transition EoS. 
The latter mixed phase region is very narrow in the considered 
range of temperatures and densities. 
Within the hadronic scenario the expansion-stage trajectories
at high energies start from considerably lower net-baryon densities 
as compared with those in the deconfinement scenarios. 
This is a consequence of the fact that the hadronic EoS is stiffer 
than the deconfinement ones at high net-baryon densities and hence 
the system is more resistant to the compression. 
The first-order transition manifests itself (in the middle panel) 
by certain irregularity of spacing between the trajectories. 
The expansion-stage trajectory at $\sqrt{s_{NN}}=$ 7.7 GeV starts below 
the phase transition border in the first-order-transition scenario. 
However, this is an effect of averaging the density 
and temperature over the whole system. Locally the system is well above 
this border at 7.7 GeV, as it was demonstrated in Ref.~\cite{Ivanov:2013wha}.

%
\begin{figure}[bht]
\includegraphics[width=8.5cm]{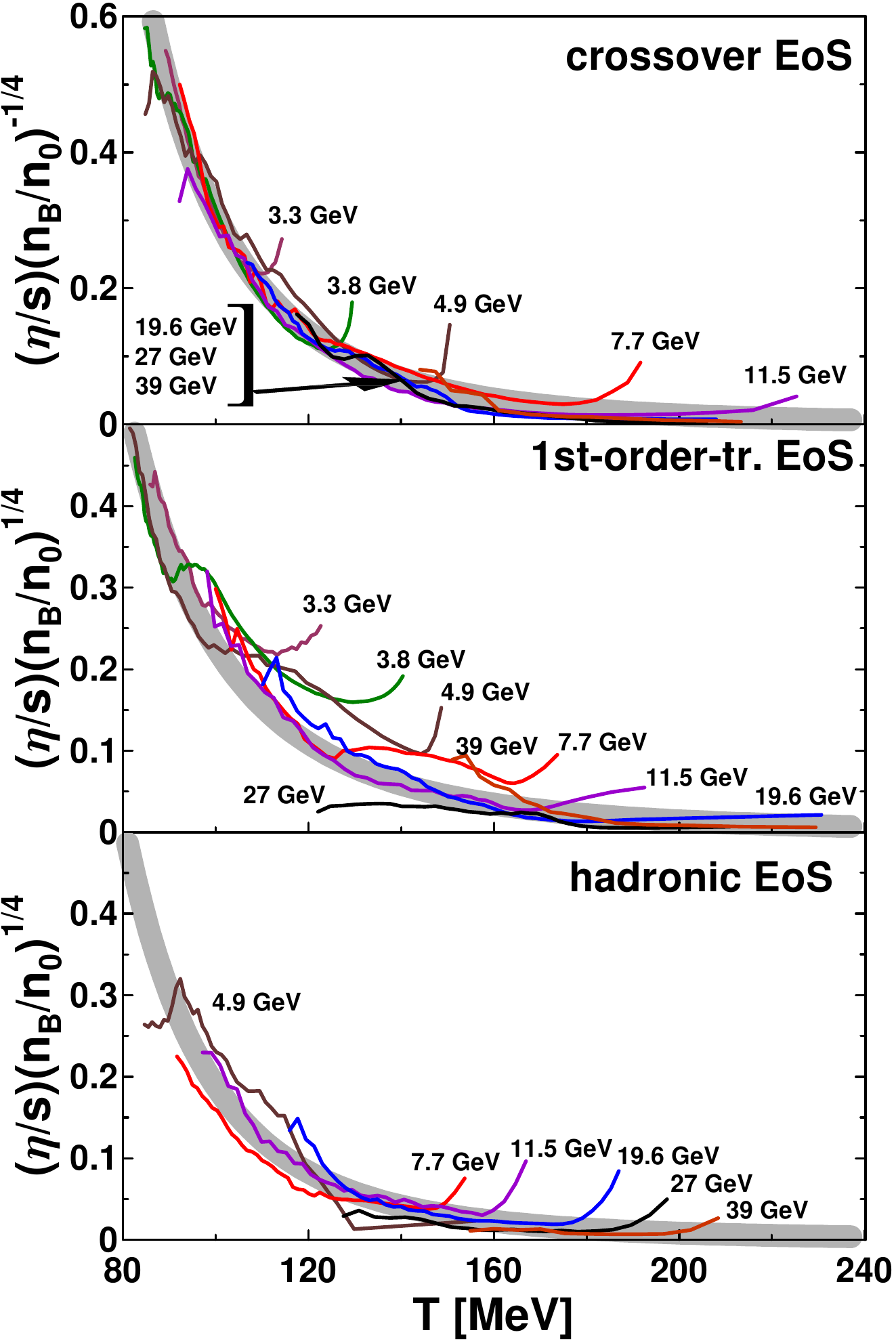}
 \caption{
The smoothed $\eta/s$ ratio scaled by powers of the proper net-baryon density ($n_B$)
as a function of temperature along trajectories of central Au+Au collisions at 
various collision energies $\sqrt{s_{NN}}$ 
within different scenarios. 
The gray bands present fits of Eqs. (\ref{eta-s-nb-T-mix})-(\ref{eta-s-nb-T-gas}).
}
\label{fig7}
\end{figure}

In order to disentangle the temperature and density dependencies of the $\eta/s$ ratio
along dynamical trajectories, we scale the $\eta/s$ ratio by powers of the proper 
net-baryon density, as it is displayed in Fig. \ref{fig7}. 
This way we tried to find a universal $T$ dependence for the $\eta/s$ ratios
at all collision energies within a single scenario. The results of such a fit 
are summarized as follows 
   \begin{eqnarray} 
   \label{eta-s-nb-T-mix}
\left( 
\frac{\eta}{s} \right)_{\rm{crossover}}  
&\approx&  \left(\frac{T_{\rm{cr.}}}{T}\right)^{4.5} \left(\frac{n_B}{n_0}\right)^{1/4}
\\ \nonumber 
&\mbox{with}& \quad T_{\rm{cr.}} = \mbox{77 MeV}, 
\\
   \label{eta-s-nb-T-tph}
\left( 
\frac{\eta}{s} \right)_{\rm{1st-order-tr.}}  
&\approx&  \left(\frac{T_{\rm{1st}}}{T}\right)^4 \left(\frac{n_B}{n_0}\right)^{-1/4}
\\ \nonumber 
&\mbox{with}& \quad T_{\rm{1st}} = \mbox{69 MeV}, 
\\
   \label{eta-s-nb-T-gas}
\left( 
\frac{\eta}{s} \right)_{\rm{hadronic}}  
&\approx&  \left(\frac{T_{\rm{had.}}}{T}\right)^{4.5} \left(\frac{n_B}{n_0}\right)^{-1/4}
\\ \nonumber 
&\mbox{with}& \quad T_{\rm{had.}} = \mbox{69 MeV}, 
   \end{eqnarray}
where $n_0=$ 0.15 fm$^{-3}$ is the normal nuclear density. 
As it is seen, the best 
result is achieved for the crossover scenario---different curves turn out to be 
well described by a universal $T$ dependence. For other scenarios the results are not 
that good. However, the spread of different curves is definitely smaller for  
the $n_B$-scaled  $\eta/s$ ratio than for those without scaling (Fig. \ref{fig4}). 
The deduced density dependences $\sim n_B^{\pm 1/4}$ are very weak, 
and the corresponding temperature dependences are very similar for all scenarios.

\section{Summary}
\label{Summary}

We calculated the entropy production in central Au+Au collisions at collision energies
from $\sqrt{s_{NN}}=$ 3.3 GeV to 39 GeV within different scenarios in order to quantify 
the dissipation in the 3FD model. 
To estimate this dissipation in therms of the effective shear viscosity (more precisely, 
the $\eta/s$ ratio), we considered this entropy as if it is generated within the 
conventional one-fluid viscous hydrodynamics. 
This effective effective shear viscosity is not a true property of  
the matter near equilibrium. It is just a  shear viscosity that would 
produce the same entropy as that resulting from the nonequilibrium self-diffusion in the 
3FD model.

It is found that more that 80\% entropy is produced during a short early 
stage of the collision which lasts $\sim$1 fm/c at highest considered energies  
 $\sqrt{s_{NN}}\gsim$ 20 GeV. 
At low collision energies $\sqrt{s_{NN}}\lsim$ 5 GeV this can be considered as an explanation of  
 the fast initial equilibration of the system because it is achieved by use of the 
 microscopically estimated friction forces in the hadronic phase \cite{Sat90}. 
At higher collision energies $\sqrt{s_{NN}}>$ 5 GeV the 3FD model 
only simulates this  fast thermalization because it is a result of purely 
phenomenological friction forces in the QGP that were tuned to reproduce 
the observed baryon stopping  \cite{Ivanov:2013wha}.

At final stages of the central Au+Au collisions 
the  $\eta/s$ ratio takes values from $\sim$0.05 at highest 
considered energies to $\sim$0.5 at lowest ones, which approximately agrees with 
earlier estimates \cite{Khvorostukhin:2010aj}. 
This result also does not contradict the finding of Ref. 
\cite{Karpenko:2015xea}, where average $\eta/s$ over the expansion stage values were reported, 
because in our case the  $\eta/s$ ratio turns out to be strongly temperature dependent.
However, 
at the initial stages of the expansion (right after the fast entropy production stage) 
in collisions at $\sqrt{s_{NN}}>$ 10 GeV
the  $\eta/s$ ratio happens to 
be noticeably smaller than the KSS bound for this quantity $1/(4\pi)$ 
\cite{Kovtun:2004de}. 
This is certainly a theoretical shortcoming of the model.

It was found that the $\eta/s$ ratio within different considered 
scenarios (with and without deconfinement transition) 
is very similar at the expansion stage of the collisions:  
\begin{itemize}
	\item 
as a function of temperature ($T$), 
$\eta/s \sim 1/T^{4\div 4.5}$ and quantitatively is very similar within different 
scenarios,  
	\item 
the $\eta/s$ ratio exhibits a weak dependence on the proper net-baryon density,  
	\item 
at final stages of the collisions 
the  $\eta/s$ ratio ranges from $\sim$0.05 at highest 
considered energies to $\sim$0.5 at lowest ones. 
\end{itemize}
Apparently, this similarity is the main reason why all considered scenarios equally well 
reproduce the measured integrated elliptic flow of charged 
particles \cite{Ivanov:2014zqa}.

\vspace*{3mm} {\bf Acknowledgments} \vspace*{2mm}

Fruitful discussions with D.N. Voskresensky 
are gratefully acknowledged.
We are grateful to A.S. Khvorostukhin, V.V. Skokov,  and V.D. Toneev for providing 
us with the tabulated 2-phase and crossover EoS's. 
The calculations were performed at the computer cluster of GSI (Darmstadt).


\begin{thebibliography}{999}
%
\bibitem{Berges:2012ks} 
  J.~Berges, J.~P.~Blaizot and F.~Gelis,
  J.\ Phys.\ G {\bf 39}, 085115 (2012)
  [arXiv:1203.2042 [hep-ph]].
%
\bibitem{Fukushima:2016xgg} 
  K.~Fukushima,
  arXiv:1603.02340 [nucl-th].
%
\bibitem{Heinz:2013th} 
  U.~Heinz and R.~Snellings,
  Ann.\ Rev.\ Nucl.\ Part.\ Sci.\  {\bf 63}, 123 (2013)
  [arXiv:1301.2826 [nucl-th]].
%
\bibitem{Kestin:2008bh} 
  G.~Kestin and U.~W.~Heinz,
  Eur.\ Phys.\ J.\ C {\bf 61}, 545 (2009) 
  [arXiv:0806.4539 [nucl-th]].
%
\bibitem{Adamczyk:2012ku} 
  L.~Adamczyk {\it et al.}  [STAR Collaboration],
  Phys.\ Rev.\ C {\bf 86}, 054908 (2012)
  [arXiv:1206.5528 [nucl-ex]].
%
\bibitem{Petersen:2008dd} 
  H.~Petersen, J.~Steinheimer, G.~Burau, M.~Bleicher and H.~Stocker,
  Phys.\ Rev.\ C {\bf 78}, 044901 (2008)
  [arXiv:0806.1695 [nucl-th]].
%
\bibitem{Karpenko:2015xea} 
  I.~A.~Karpenko, P.~Huovinen, H.~Petersen and M.~Bleicher,
  Phys.\ Rev.\ C {\bf 91}, no. 6, 064901 (2015)
  [arXiv:1502.01978 [nucl-th]].
%
\bibitem{Itakura:2007mx} 
  K.~Itakura, O.~Morimatsu and H.~Otomo,
  Phys.\ Rev.\ D {\bf 77}, 014014 (2008)
  [arXiv:0711.1034 [hep-ph]].
%
\bibitem{Khvorostukhin:2010aj} 
  A.~S.~Khvorostukhin, V.~D.~Toneev and D.~N.~Voskresensky,
  Nucl.\ Phys.\ A {\bf 845}, 106 (2010)
  [arXiv:1003.3531 [nucl-th]].
%
\bibitem{Denicol:2013nua} 
  G.~S.~Denicol, C.~Gale, S.~Jeon and J.~Noronha,
  Phys.\ Rev.\ C {\bf 88}, no. 6, 064901 (2013)
  [arXiv:1308.1923 [nucl-th]].
%
\bibitem{Kadam:2015xsa} 
  G.~P.~Kadam and H.~Mishra,
  Phys.\ Rev.\ C {\bf 92}, no. 3, 035203 (2015)
  [arXiv:1506.04613 [hep-ph]].
%
%
\bibitem{Ivanov:2014zqa} 
  Y.~B.~Ivanov and A.~A.~Soldatov,
  Phys.\ Rev.\ C {\bf 91}, no. 2, 024914 (2015)
  [arXiv:1401.2265 [nucl-th]].
%
\bibitem{3FD}
 Yu. B. Ivanov, V. N. Russkikh, and V.D. Toneev,
 Phys. Rev. C {\bf 73}, 044904 (2006) 
[nucl-th/0503088].
%
%
\bibitem{gasEOS}
V. M. Galitsky and I. N. Mishustin, Sov. J. Nucl. Phys. {\bf 29}, 181 (1979).
%
%
\bibitem{Toneev06}
A. S. Khvorostukhin,  
V. V. Skokov, K. Redlich, and V. D. Toneev,
Eur. Phys. J. {\bf C48}, 531 (2006)  [nucl-th/0605069].
%
\bibitem{FOPI05}
A. Andronic {\it et al.} (FOPI Collaboration), Phys. Lett. {\bf B612},
173 (2005) 
[arXiv:nucl-ex/0411024].
%
\bibitem{Ivanov:2016vkw} 
  Y.~B.~Ivanov and A.~A.~Soldatov,
  Eur.\ Phys.\ J.\ A {\bf 52}, no. 5, 117 (2016)
  [arXiv:1604.03261 [nucl-th]].
%
%
{
\bibitem{Land-Lif}
L. D. Landau  and E. M. Lifshitz, {\em Fluid Mechanics} (Pergamon
Press, Oxford, 1987).  
}
%
\bibitem{Rischke:1998fq} 
  D.~H.~Rischke,
  Lect.\ Notes Phys.\  {\bf 516}, 21 (1999)
  [nucl-th/9809044].
%
\bibitem{Ivanov:2013wha} 
  Yu.~B.~Ivanov,
  Phys. Rev. C {\bf 87}, 064904 (2013) [arXiv:1302.5766 [nucl-th]]. 
%
\bibitem{Ivanov:2013yqa} 
  Yu.~B.~Ivanov,
  Phys. Rev. C {\bf 87}, 064905 (2013) [arXiv:1304.1638 [nucl-th]]. 
%
%
\bibitem{Russkikh:2006aa} 
  V.~N.~Russkikh and Yu.~B.~Ivanov,
  Phys.\ Rev.\ C {\bf 76}, 054907 (2007)  [nucl-th/0611094];
%
\bibitem{Ivanov:2008zi} 
  Yu.~B.~Ivanov and V.~N.~Russkikh,
  Phys.\ Atom.\ Nucl.\  {\bf 72}, 1238 (2009)  [arXiv:0810.2262 [nucl-th]].
%
\bibitem{Horvat:2010dk} 
  S.~Horvat, V.~K.~Magas, D.~D.~Strottman and L.~P.~Csernai,
  Phys.\ Lett.\ B {\bf 692}, 277 (2010)
  [arXiv:1007.4754 [nucl-th]].
%
\bibitem{Csernai:2011qq} 
  L.~P.~Csernai, D.~D.~Strottman and C.~Anderlik,
  Phys.\ Rev.\ C {\bf 85}, 054901 (2012).
%
\bibitem{Brac98}
M.~Reiter, A.~Dumitru, J.~Brachmann, J.A.~Maruhn, H.~St\"ocker,
and W.~Greiner, Nucl. Phys. {\bf A643}, 99 (1998) [nucl-th/9806010].
%
\bibitem{Kovtun:2004de} 
  P.~Kovtun, D.~T.~Son and A.~O.~Starinets,
  Phys.\ Rev.\ Lett.\  {\bf 94}, 111601 (2005)
  [hep-th/0405231].
%
\bibitem{Sat90} L. M.~Satarov, Yad. Fiz. {\bf 52}, 412 (1990)
[Sov. J. Nucl. Phys. {\bf 52}, 264 (1990)].


\end{thebibliography}
\end{document}